\documentclass{emulateapj}
\usepackage{ulem}

\slugcomment{}

\shorttitle{Fluorine and Sodium in C-rich low-metallicity stars}
\shortauthors{Lucatello et al.}

\begin{document}


\title{Fluorine and Sodium in C-rich low-metallicity stars \altaffilmark{1,2}}


\author{Sara Lucatello \altaffilmark{3,4}
Thomas Masseron \altaffilmark{5}, Jennifer A. Johnson \altaffilmark{5}, Marco Pignatari \altaffilmark{6,7,8}, 
Falk Herwig \altaffilmark{6}}
\altaffiltext{1}{Based on observations made with ESO Telescopes at Paranal Observatories 
under programme ID  080.D-0606A}
\altaffiltext{2}{This publication makes use of data products from the Two Micron All Sky
    Survey, which is a joint project of the University of Massachusetts and the
    Infrared Processing and Analysis Center, funded by the National Aeronautics
    and Space Administration and the National Science Foundation}
\altaffiltext{3}{INAF--Osservatorio Astronomico di Padova, vicolo dell'Osservatorio 5, 35122,
Padova, Italy}
\altaffiltext{4}{Excellence Cluster Universe, Technische Universit\"at M\"unchen, 
Boltzmannstr. 2, D-85748, Garching, Germany}
\altaffiltext{5}{Astronomy department, Ohio State University, 
140 W. 18th Ave., Columbus, OH 43210; jaj@astronomy.ohio-state.edu, 
masseron@astronomy.ohio-state.edu}
\altaffiltext{6}{Department of Physics \& Astronomy, University of Victoria, Victoria, 
BC V8P5C2, Canada}
\altaffiltext{7}{Joint Institute for Nuclear Astrophysics, University of Notre Dame, IN
46556, USA}
\altaffiltext{8}{TRIUMF, 4004 Wesbrook Mall, Vancouver, BC, Canada, V6T 2A3}

\begin{abstract}
We present the N, O, F and Na abundance and $^{12}$C/$^{13}$C isotopic ratio measurements 
or upper limits for a sample of 10 C-rich, metal-poor giant stars,
eight enhanced in s-process (CEMP-s) elements and two poor in n-capture elements
(CEMP-no).
The abundances are derived from IR, K-band, high-resolution CRIRES@VLT spectra obtained.
 The metallicity of our sample ranges from [Fe/H]=$-$3.4 to $-$1.3. 
F abundance could be measured only in two CEMP-s stars. With [F/Fe]$=$0.64, one is mildly F-overabundant, while the other is F-rich, at [F/Fe]$=$1.44. 
For the remaining eight objects, including both CEMP-no in our sample, only upper limits on F abundance 
could be placed. Our measurements and upper limits show that there
is a spread in [F/C+N] ratio in CEMP-s stars
as predicted by theory.
Predictions from nucleosynthetic models for 
low-mass, low-metallicity Asymptotic Giant Branch stars, account for  the derived F abundances, while the upper limits on F content derived for most of the stars  
are lower than the predicted values.
The measured Na content is accounted for 
by AGB models in the 1.25 to 1.75\,M$_{\odot}$ range, confirming that 
the stars responsible for the peculiar abundance pattern observed in 
CEMP-s stars are low-mass, low-metallicity AGB stars, in agreement 
with the most  accepted astrophysical scenario. 
We conclude that the mechanism of F production in 
current state-of-the-art low-metallicity low-mass AGB models 
needs further scrutiny and that F measurements in a larger number of metal-poor stars are needed to better constraint the models.
\end{abstract}





\section{Introduction}
The understanding of the stellar nuclear production sites and evolution scenarios
 is of great current interest. A large number of high-resolution spectral studies 
targeting metal-poor 
objects selected by mining the HK survey \citep{beers92} and the HES 
\citep{christlieb01} have measured a wide set of chemical elements
ranging from Li \citep[e.g.][]{spite05} to the
n-capture species \citep[e.g.][]{sneden03}, providing
constraints on how the stars formed as well as on the nucleosynthetic
processes that took place in the early Galaxy. 

In contrast, only one single
measurement of F in extremely metal-poor (EMP) stars has been published to date \citep{schuler07}.
The most accessible lines suited for F measurements, the vibration-rotation 
transition of the HF molecule, are located in the K band, requiring the use of 
high-resolution, IR spectrographs mounted on 8\,m class telescope to target halo stars, 
a configuration not available until recently.

Fluorine is an element of  particular interest, extremely sensitive to
the physical conditions within stars. Although the single stable F isotope, 
$^{19}$F, is not involved in the main reactions taking place in the cores of 
stars, it can be created and destroyed in several
different ways, and its dominant source is not yet clear.
Theoretical modeling has indicated (at least) three possible sites for F
production.

\citet{woosley98} proposed first that F is produced by neutrino spallation
on $^{20}$Ne; $^{20}$Ne($\nu$,$\nu ' p$)$^{19}$F.  
Second, modeling  shows that low-
metallicity, low-mass (M$\leq$3-4\,M$_{\odot}$) AGB stars synthesize very 
large quantities of F \citep[see e.g.][]{cristallo07} via the 
reactions $^{14}$N(n,p)$^{14}$C($\alpha$,$\gamma$)$^{18}$O($p,\alpha$)$^{15}$N($\alpha,\gamma$)$^{19}$F {\bf and 
$^{14}$N($\alpha$,$\gamma$)$^{18}$F($\beta^{+}$)$^{18}$O(p,$\alpha$)$^{15}$N($\alpha$,$\gamma$)$^{19}$F},  
where the neutrons are provided by $^{13}$C($\alpha,n$)$^{16}$O
and the protons mainly by $^{14}$N($n,p$)$^{14}$C \citep[see][]{jorissen92}.
On the other hand, 
massive AGB stars are expected to destroy F via Hot Bottom Burning (HBB) 
\citep[see e.g.][]{smith05,karakas07}. 
Finally, \citep{meynet00} proposed that Wolf-Rayet stars are a possible source for F, through the $^{14}$N($\alpha,\gamma$)$^{18}$
F($\beta^{+}$)$^{18}$O($p,\alpha$)$^{15}$N($\alpha,\gamma$)$^{19}$F  chain.
While the physical conditions for F production exist in several phases of 
stellar evolution, the precise contribution of each of these three sites to the 
Galactic F evolution is not known, 

The measurements of F content in any object have been few and give
conflicting evidence about the origin of F in the Universe.
\citet{renda04} showed that the inclusion of all of sites in Galactic
chemical evolution models is necessary to reproduce the F abundances
measured in Milky Way field stars. In their model, AGB stars are
responsible for significant amounts of F production in the early
Universe, because of the metallicity dependence of their yields, while
WR stars are significant contributors at solar and super-solar
metallicities.  \citet{jorissen92} derived F abundances in giants that
recorded the chemical evolution (K and M giants), in AGB giants that
were dredging up freshly minted F from their interiors (MS, S, SC, N
and J giants), and in giants that had been polluted by one of
the former (Ba giants). They suggested that the overproduction of F in
the AGB stars (factors of 3 to 30 with respect to the solar system in
the most extreme S and N stars) showed that AGB stars are likely to
major contributors to the Galactic F abundance, at least at
metallicities close to solar.  In addition, the largest F
overabundances measured could not be explained with standard AGB
models and required additional mixing to achieve the desired amounts.
However, recent results by \citet{abia08} and \citet{abia10} suggest
that because of a possible lack of proper accounting for C-bearing
molecules (i.e. CH, CN, CO and C$_2$) contribution, the F abundances
reported in \citet{jorissen92} for solar metallicity giants, which
relied on the same HF line used in the present analysis, had 
been overestimated.  Therefore, the long standing problem of the
discrepancy between the high [F/Fe]\footnote{
Hereafter:$\log \epsilon$(A) = log N(A)+12.  [X/Y] =
log(X/X$_{\odot}$) $-$ log(Y/Y$_{\odot}$).} and [F/O] ratios measured in solar
metallicity giants and the low ones resulting from the models, is
likely solved by the adoption of a more complete molecular line list
for the synthetic spectra.  In fact, the \citet{cristallo07} models
account fully for the F abundances measured in several C-stars ranging
from solar metallicity to slightly metal-poor ([Fe/H]$\sim$0.0 to
$\sim$-0.5) \citep{abia08,abia10}.

While the above results make it clear that solar metallicity AGB stars
make at least some F, there is not yet a consensus on the overall
contribution of AGB stars throughout the chemical evolution of stellar
systems. \citet{cunha03} measured F in red giants in the LMC and
$\omega$ Cen. They found that F/O declines as the O abundance
decreases and the two $\omega$ Cen giants have particularly low F/O
values. They argue that their results are qualitatively consistent
with most F production coming from either neutrino
nucleosynthesis or WR stars rather than AGB stars in these systems,
which certainly have a different chemical evolution than the Milky
Way. In particular, the low F abundances in metal-poor
([Fe/H]$\sim -1$) $\omega$ Cen stars that are enriched in s-process
elements is difficult to reconcile with the idea that metal-poor AGB
stars are major contributors to F production.  F has also been
measured in stars in the Galactic bulge \citep{cunha08}.  The F/O
results there can be explained by contributions by both AGB and WR
stars, although the lack of an s-process enhancement in the most
F-rich bulge stars suggests that AGB stars may play a less prominent
role in the bulge than has been inferred for the disk.

Measurements of fluorine in the interstellar medium \citep{federman05}
 show no evidence of F over-abundances due to the neutrino process in
 SNII.  The F production in rotating WR stars has been reconsidered by
 \citet{palacios05}, who found that F yields are significantly lower
 than the \citet{meynet00} predictions, indicating that their
 contribution to the Galactic F budget would be negligible. These
 results suggest that at low-metallicity AGB stars play a major role
 in F production, although this idea would leave the $\omega$ Cen
 results unexplained.  The idea of AGB stars as producers of F is
 supported by the large F enhancements found in post-AGB stars
 \citep{werner05} and planetary nebulae \citep[see e.g.][and
 references therein]{otsuka08}, the progeny of AGB stars as well as
 the results for abundances in Milky Way AGB stars cited above. However, these
observations are focused on [Fe/H]$\sim$0 metallicity stars and do not
address directly the question of the production of F production in
metal-poor AGB stars.

While Na measurements have been obtained for a much larger number of 
metal-poor stars, no quantitative study has been performed so 
far on the role of AGB stars in its production, in particular at low metallicity.
In thermally pulsing AGB stars, Na is mostly
synthesized by proton captures on $^{22}$Ne.
Models indicate that the main source of Na in low-mass AGB stars, 
at least at solar metallicity,  is the creation of a 
Na pocket, located at the top (or near the top) of the $^{14}$N 
pocket. At lower metallicities, other mechanisms, 
such as the neutron capture on $^{22}$Ne,
which can occur both during the radiative $^{13}$C
burning and during the convective $^{22}$Ne
burning, may become important (Cristallo et al 2006b and references therein). 
In general, according to e.g., Cristallo et al. and \citet{bisterzo10}, the production of Na increase by about 1\,dex or more going from solar to [Fe/H] ~ -2.3, a similar metallicity to that of the stars in our sample. This is due to a higher efficiency of all the Na nucleosynthesis channels described above.
The testing of such predictions is important to constraint details of the key reactions involved.

Carbon-enhanced metal-poor (CEMP) stars provide an opportunity to directly
measure the F and Na production in low-mass, metal-poor AGB stars. These
stars are chemically peculiar
objects, characterized by an overabundance of C ([C/Fe]$>$1\footnote{
Some authors use different cutoffs values depending on the evolutionary state 
of the star, 
with cutoff enhancement for giants as low as 
[C/Fe]$\simeq$0.5 \citep[see][]{aoki07}}) accounting for
10-20\% of stars below [Fe/H]$\leq-$2.5
\citep{marsteller05,cohen05,lucatello06}.  Less than a third of CEMP
stars exhibit no enhancement in heavy elements (CEMP-no), while most
of these objects \citep[over 70\% ][]{aoki07} are characterized, by
an overabundance of n-capture, s-process, elements (CEMP-s).

\citet{lucatello05} showed that likely {\it all}
C-rich, extremely metal-poor stars with s-process enhancement (the
CEMP-s stars) belong to a binary system. CEMP-s are then the metal-poor 
analog to the classical CH and Ba stars: low-mass stars (M$\sim$0.8\,M$_{\odot}$)
whose slightly more massive  (between $\simeq$1.2 and $\simeq$2.5\,M$_{\odot}$, the exact 
range depending on metallicity)
 companion, now a faint white dwarf, dumped material processed during
 its AGB phase on their surfaces, leaving its chemical fingerprints in
 the composition of their envelopes. Therefore, the nucleosynthetic
 processes taking place in extremely metal-poor, low-mass ($\simeq$1.5\,M$_{\odot}$)
 stars, now long extinct, can be investigated through the study of
 CEMP-s stars characteristics \footnote{Because of its high Eu abundance, 
not fully accountable with 
standard s-process nucleosynthesis, one of the stars in our sample, 
HD187861, has been included in the CEMP-rs category. However, since all
 the likely formation scenarios invoked to explain 
the abundance patterns of these objects include mass transfer from an 
AGB companion \citep[see e.g.][]{jonsell06,lucatello09}, as far as the 
present discussion is concerned HD187861 can be considered as an CEMP-s}.

The origin of CEMP-no objects is, on the other hand, still a mystery.
\citet{fujimoto00} suggested that they may have formed as chemically
``normal'' low-mass stars and became C-enhanced through a path of
self-enrichment due to anomalous mixing processes specific to
low-metallicity stars. Alternatively, as  \citet{ryan05} suggested, 
they could have been born from C-rich gas, possibly polluted by a previous 
generation of supernovae whose fall-back avoids the ejection of heavier elements
during the explosion  \citep[e.g.,from high-mass, rapidly rotating stars see][and references therein]{meynet06}. Objects like the recently discovered extremely-metal poor, C-rich Damped Lyman-$\alpha$ system \citep{cooke10} might turn out to be the connection between the yields of 
the Pop III stars and their later incorporation into CEMP-no stars.
   
On the other hand, their abundance patterns could arise from early AGB transfers from 
low-mass stars before any considerable s-process element production took place
\citep{ryan05,masseron09a}, or alternatively  from an AGB stars whose evolution 
was truncated by binary interaction (see, e.g., Izzard \& Tout 2003)
with its companion, the presently observed object.

The measurement of a value or a strong upper limit on F abundance is of crucial importance to probe the origin of  the chemical pattern observed in CEMP-no stars given that F can be
synthesized before the bulk of the s-process-element production.
In \citet{cristallo07}, for instance, as early as the third-dredge-up episode 
for a 2\,M$_{\odot}$ at [Fe/H]=$-$2.3, [F/C]=-0.4 and [F/Fe]=1.65, [Ba/C]=-1.5 and [Ba/Fe]=0.5 (and [C/Fe]=2.05).
Therefore, a measurement in CEMP-no stars of an amount of F comparable to C would
strongly argue in favor of AGB enrichment.

\citet{schuler07} measured the F abundance in a extremely 
metal-poor star for the first time, deriving 
an abundance of
 $\log \epsilon$(F)= $+4.96 \pm 0.21$ corresponding to an abundance ratio 
[F/Fe] = $+$2.9 for the CEMP-s star HE~1305+0132. \citet{lugaro08} 
compared this value to existing nucleosynthesis and mass transfer models. 
Conclusion that an object with such an extreme 
F content should be exceedingly rare, while most CEMP-s stars are expected to
 exhibit a noticeable but smaller F overabundance.

We here present F abundance measurements for two CEMP-s and upper limits
for eight more CEMP stars (six CEMP-s and two CEMP-no) and discuss the
implications of our result on our current understanding of AGB
nucleosynthesis. 

\section{Observations} 
Data were collected using
CRIRES\footnote{www.eso.org/instruments/crires/}
 (CRyogenic high-resolution InfraRed Echelle Spectrograph,
 \citealt{kaeufl04}) at VLT UT1 (Antu) in a series of service runs
 between October 21th 2007 and February 28th 2008.  The slit-width was
 set to 0\farcs4 yielding a resolution of
 $R=\lambda/\Delta\lambda\simeq50,000$, where $\Delta\lambda$ refers
 to the width of the resolution elements, with a binning of
 1$\times$1.  We used the 24/-i/i setup, which leads a wavelength
 coverage from 2299.3 to 2311.8\,nm, 2315.2 to 2327.3\,nm, 2330.4 to
 2342.0\,nm, and 2345.0 to 2356.0\,nm\ for chips $\#$1 to 4
 respectively.  Our sample also included a well-studied C-normal metal-poor star (HD~122563) to be used as a comparison star.  The typical
 signal-to-noise ratio for the program stars at $\sim$2335\,nm is
 about 120 per pixel.  An early-type, fast-rotating star was
 observed right before or right after each one of the program star
 observations.  To improve the accuracy of the telluric line
 subtraction procedure, those objects were chosen so that the science
 and calibration stars were observed at similar air-masses.  The
 typical signal-to-noise ratio for the telluric standard stars at
 $\sim$2335\,nm\ is about 250 per pixel.

The raw frames were processed with the CRIRES GASGANO pipeline
(version 1.6.0). The 1D science and standard star spectra were
wavelength-calibrated separately using the numerous telluric
absorption lines present on all of the four detector arrays.  Such
process was performed separately for the science and telluric standard
star spectrum because of the limited reproducibility of the Echelle
grating position.  The science spectra were then divided by the
appropriate standard star spectra to correct for the telluric lines
and the illumination pattern in the best possible way.  It is worth
noting that while the telluric lines are strong enough to be used
 for wavelength calibration, they are weak enough to be accounted for by
standard star division; hence their effect on the abundance measurements 
presented in this paper is negligible.

More details about the observations can be found in Table \ref{tab1}.



\section{Atmospheric parameters and atmospheric models} \label{sec_atm}

The determination of the adopted atmospheric parameters can be obtained
photometrically, via the derivation of the effective temperature
T$_{\rm eff}$ from accurate T$_{\rm eff}$-color relations and that of surface
gravity, $\log g$ from a suitable isochrone.  Alternatively, T$_{\rm eff}$
and $\log g$ can be derived spectroscopically, through the requirement
of, respectively, of excitation and ionization equilibrium.  In both
cases, the micro turbulent velocity $\xi$ is determined from
minimizing the trend of derived abundance and equivalent widths of the
Fe lines.

The spectral range covered by the program spectra do not include any
Fe lines and in general not enough features to allow the derivation of
the atmospheric parameters. Therefore, we have relied on the
atmospheric parameters published in the recent literature or derived
on the basis of photometry and our own proprietary optical spectra
(for details see Masseron et al. 2010a).  The adopted atmospheric
parameters and C and Ba abundances (which cannot be 
derived from our spectra as no suitable features are present in the 
$K$ band), along with their uncertainties and sources are listed in Table \ref{tab1}.
Note that in some cases the uncertainties are not listed as the
source paper did not report them.

The chemical composition of the stellar atmosphere can affect its
structure: the higher the content of metals the larger the number of
free electrons, available for the formation of the H$^-$ ion, the main
source of continuum opacity among stars with spectral types later than
A.  Solar-scaled model atmospheres, which are generally
(appropriately) adopted for the analysis of stars, likely provide
different temperature profiles from those that are expected to be
found in CEMP stars, due to their large C and N enhancements.  In our
abundance analysis, we adopt C and N enhanced model atmospheres belonging to
the OSMARCS family \citep{gustafsson08} calculated using the
appropriate atmospheric parameters and C and N enhancements for each star.
Such models have a steeper temperature profile than their solar-scaled
counterparts, affecting especially the computation of the strong
lines, which form in the outer layers.  The discussion of the
calculation of the C and N-enhanced models, their features and their impact
on the abundance determinations is beyond the aim of the present
paper.  For more details the reader is referred to \citet{masseron09}.

\section{Abundance analysis and comparison with the literature.}\label{sec_abundance}

Elemental abundances for N, O, F, Na and C isotopic ratios 
were obtained by synthetic spectra
fits to the 1D extracted spectra for the sample stars.  We used the
TURBOSPEC synthesis code \citep{alvarez98}, which
shares routines and input data with OSMARCS. The code is characterized
by chemical equilibrium including 92 atomic and over 500 molecular
species, and Van der Waals collisional broadening by H, He, and H$_2$
following \citet{barklem05}.
{\bf Smoothing for the spectra was determining by simultaneous fitting of a number 
of weak unblended CO lines.

A partial analysis of this dataset has been presented in \citet{lucatello09}. The present 
paper presents a reanalysis of the data, providing measurements for a wider set of elements. 
It is based on improved reduction of the spectra (using a different 
pipeline version) and  makes use of more accurate atmospheric parameters determinations. Hence the 
present results supersede those presented in \citet{lucatello09}. 
}


O was measured from the fit of several $^{12}$C$^{16}$O lines
belonging to the $\nu$ = 2-0 and 3-1 overtone bands, as well as
 the $\nu$ = 4-2 and $^{13}$CO $\nu$ = 2-0 band-head regions. 
Given the high C content of the stars under consideration, the CO lines 
are very sensitive to O content, hence allowing for a very accurate O 
abundance measurement. 
$^{12}$C/$^{13}$C isotopic ratios were measured from the simultaneous fitting 
of several $^{12}$C$^{16}$O and $^{13}$C$^{16}$O lines. 
O abundance was then re-measured with the appropriate C isotopic ratio and the 
procedure was repeated to convergence. N abundance measurements 
are obtained from a number of CN lines scattered 
throughout our spectral range.

Three Na lines are included within the available spectral region: 2334.842\,nm,
2337.895\,nm and 2337.914\,nm. The latter is the strongest and, unlike the other 
two, is not severely blended with CO features, making it the best candidate 
for Na abundance measurements. It falls in a range of strong telluric contamination, 
therefore, in spite of careful and accurate modeling and subtraction of telluric features, 
 in four out of eleven (10 CEMP and one EMP) stars only upper limits (1$\sigma$) to the Na abundances could be placed.

\begin{figure}
\epsscale{.65}
\plotone{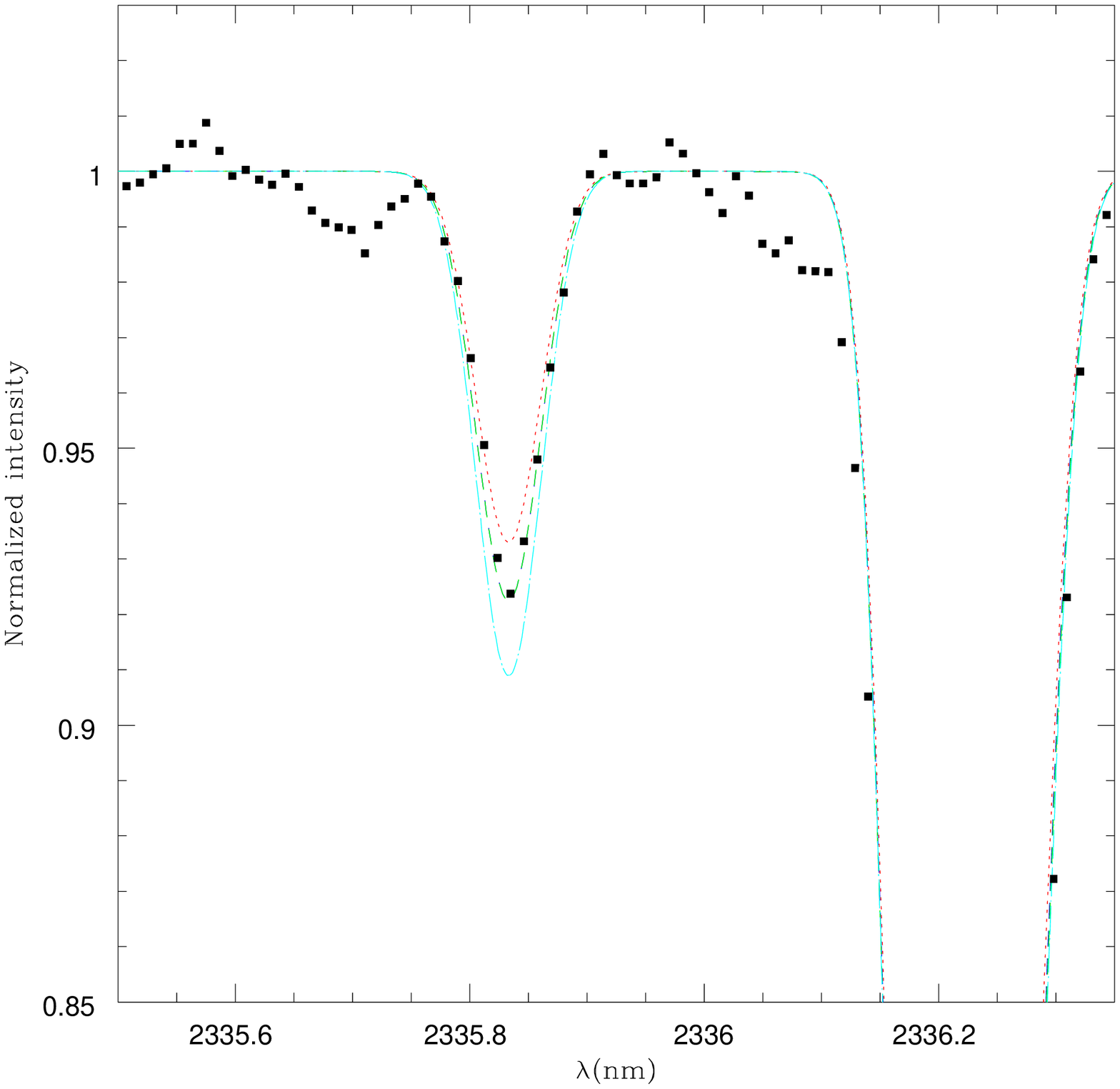}
\caption{HF line at 2335.8\,nm in star HD~5223 with synthetic 
spectra for $\log \epsilon $(F)=3.95, 4.0 and 4.05. 
\label{fig0}}
\end{figure}

Finally, F abundances (and 1$\sigma$ upper limits) were measured from the
fitting of the unblended HF line at 2335.833\,nm.  Two more HF lines
are present in our spectral range, 2324.062 and 2348.805\,nm; however
both of them are blended with CO features, hampering the accuracy of F
abundance determinations obtained from their fitting.
Figure \ref{fig0} shows the fit of synthetic spectra to the HF feature at 
2335.833\,nm.

The line list adopted are from \citet{goorvitch94} for CO, Kurucz
(1995) for CN and Na, and \citet{decin00} for HF.  They were checked for
consistency by fitting the Arcturus spectrum and that for our
comparison star HD\,122563.

In a few cases (CS\,29502-92, CS\,30314-67 and HD\,122563) no O abundance in conjunction
with the literature C abundance could yield satisfactory fits of the
CO features. Hence we sligthly adjusted the C abundance to
improve the fit of such features.  This discrepancy could be due to
the adoption of different model atmospheres in the present analysis 
and in the literature studies in which the adopted C-abundances are reported. 
In particular the
inclusion of the appropriate opacities for the C and N enhanced 
atmospheres can result into the
derivation of slightly different abundances from those obtained with
solar-scaled models, which are generally adopted in the literature.

The uncertainties in our measurements come from
two sources (excluding systematic NLTE and 3D atmosphere
effects, and atomic and molecular parameters uncertainties):
 error in the fitting of the spectral features,
and due to the atmospheric parameters.

In the case of the fitting uncertainty, we estimate this
component of the error by how well fitted we can match the
synthesis to the actual spectrum (we adopt $\sim$0.1\,dex)
To estimate the errors due to the uncertainty in the 
atmospheric parameters for each element, we determined 
individually the sensitivities of the 
derived abundances to the adopted T$_{eff}$, 
surface gravity $\log g$, C enhancement,
 and [Fe/H], and then summing in quadrature the resulting 
uncertainties associated with each parameter\footnote{Strictly speaking 
this is only appropriate in case of independent errors, and in general 
one needs to take into account the covariances. Given that the atmospheric parameters 
for most of the objects in our sample are taken from literature sources, the 
determination of the covariance terms is essentially impossible. 
On the basis of the data for the three stars for which 
atmospheric parameters were derived from our proprietary high-resolution
spectra, we obtained estimate the variations in $\log g$ and [Fe/H] for
an increase of 200\,K in T$_{eff}$, finding an average of 
~0.5\,dex and 0.15\,dex respectively.
Hence, T$_{eff}$ is clearly the dominant source of uncertainty (see Table \ref{tab3}),
and the covariance terms can be neglected.}. 
Note that 
for the stars for which no estimates for the atmospheric parameter 
uncertainties were provided in the source paper, we adopted 
100\,K for T$_{eff}$, 0.3 in $\log g$, 0.15 in [Fe/H] and 0.2 in [C/Fe].
The sensitivities to 
the adopted atmospheric parameters 
are provided in Table \ref{tab3} for HE\,1152-0355 and HD\,5223,
which beside being the only two stars for which F is measured, 
are representative of the present sample for atmospheric parameters.
The estimates of the errors reported are for abundances (e.g. 
$\log \epsilon$(F) or [F/H]), however the abundance ratios between two species (e.g. [F/Fe]) have
typically smaller errors since the effects due to uncertainties on the 
atmospheric parameters partially cancel out. As above mentioned, we cannot 
obtain the information about Fe sensitivity to atmospheric parameters in 
our sample stars, therefore we cannot actually calculate the errors on the 
abundance ratios. In the following 
plots and discussion we adopt for the ratios of F, C, N, O and Na 
to Fe the same errors, keeping in mind that they are overestimates 
of the actual ones.

The resulting N, O, F  and Na abundances, along with the observational errors, 
are listed in Table \ref{tab4}.

%
The HF lines in the spectrum of the comparison, C-normal star
HD\,122563 were undetectable as expected, thus only an upper limit
could be placed to the F abundance. This upper limit is consistent 
with the galactic chemical evolution model for such element,
which predicts a value of [F/O]$\simeq-$0.6\,dex at $\log \epsilon$(O)=7.0 
\citep[the minimum metallicity 
taken into account, see][]{renda04}, while we find [F/O]$<-$0.1\,dex at 
$\log \epsilon$(O)=6.8.

\begin{figure}
\epsscale{.65}
\plotone{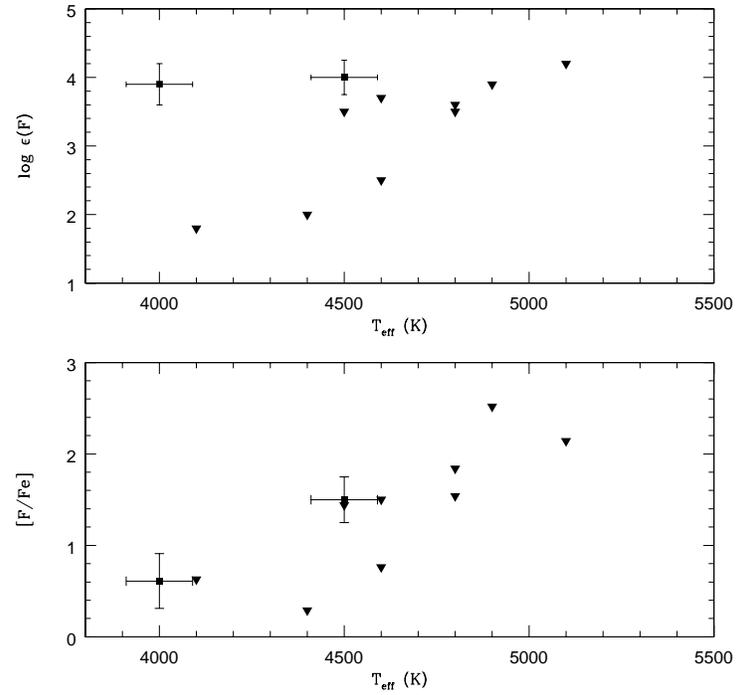}
\caption{Measured F abundance as a function of T$_{\rm eff}$ in the sample stars. Inverted triangles indicate upper limits. 
 There is a clear correlation between the upper limits set on the F abundances and 
the T$_{\rm eff}$, indicating that the most stringent limits can be obtained from cooler stars. \label{fteff}}
\end{figure}

For the program stars, as shown in Table \ref{tab3}, F
measurements could be derived only for two out of ten program stars;
with metallicities of [Fe/H]=$-$1.2 and $-$2.0
 they are the object with the highest Fe content of the sample.  For
 the other eight, more metal-poor objects, only upper limits could be
 placed.  Because of the high sensitivity of the HF molecule to
 temperature, as shown in Fig. \ref{fteff}, the level of upper limits on
 the F abundance is essentially set by the T$_{\rm eff}$ in
 the atmosphere of the observed stars.

\begin{figure}
\epsscale{.65}
\plotone{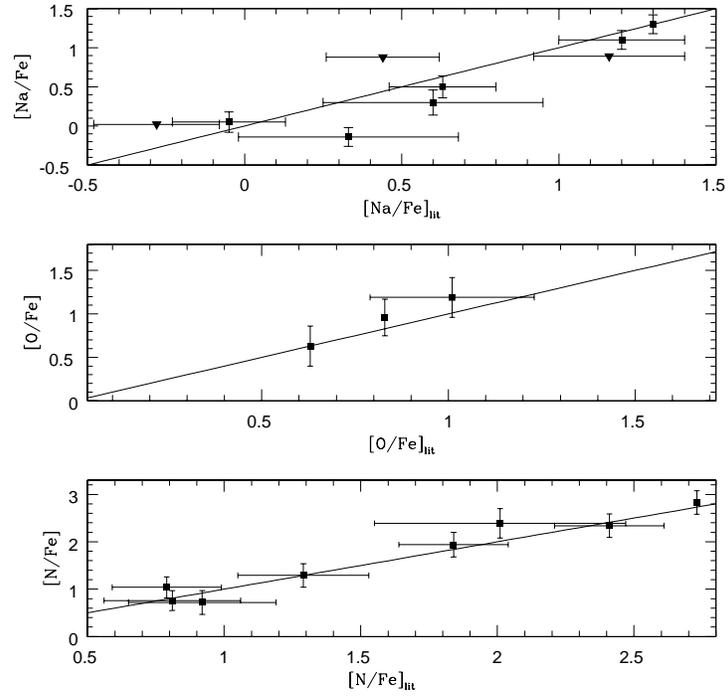}
\caption{Comparison of the derived abundances and literature for N, O and Na. Sources for literature data as in Table \ref{tab2} 
\label{flit}}
\end{figure}

As F has been measured only in one (not part of our sample) 
extremely metal-poor star before \citep[see][]{schuler07},
our results for F abundances cannot be compared to literature ones. However
measurements for N, O and Na have been previously reported for a subset of 
our sample. Fig. \ref{flit} shows the excellent agreement 
with literature results for these species, which are all measured from transitions in the 
visible spectra.  

\section{Discussion} \label{discussion}

With measurements or upper limits for eight stars, we
find a range in [F/Fe] and [F/C+N] ratios in CEMP stars (see e.g. Fig. \ref{fteff}),
with CEMP-s [F/Fe] values ranging from $+$1.44\,dex to less than $+$0.6\,dex. This
confirms both that AGB stars are characterized by
 range of efficiency in its production (likely due to different metallicity and 
and mass), as predicted by theoretical low-mass, low-metallicity nucleosynthetic
 AGB stars models. 
The two stars with measured F, HD 5223 and HE 1152-0355,
are both CEMP-s stars, and the presence of both F and s-process elements
in their atmospheres is consistent with a picture where they were
polluted by a low-mass AGB star.

There is no object in our sample with a F content as high (or close) to that
presented in \cite{schuler07} for HE 1305$+$0132. In fact, even the
highest value for an upper limit in the present sample, [F/Fe]$<$2.5
is almost a factor of three lower than those measured in HE
1305$+$0132.  
 Possible sources of systematic offsets include difference in the model
atmospheres (we use spherical C-enhanced models, whereas Schuler et
al. adopt C-enhanced, plane parallel models) and in the molecular
transition data.

\begin{figure}
\epsscale{.65}
\plotone{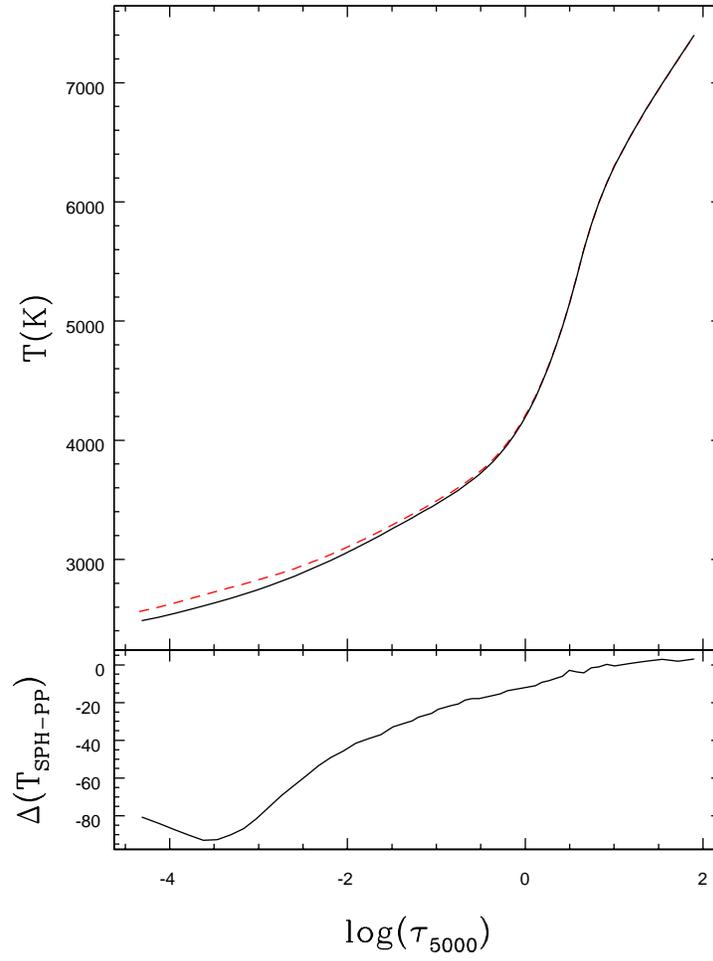}
\caption{Example of comparison between a spherical model (solid black line) and a plane-parallel one (red dashed line). The models are computed for the atmospheric parameters of HE 1152-0355 
\label{fmodels}}
\end{figure}

In our analysis we used spherical C-enhanced models. These are cooler
at the surface than the C-enhanced plane parallel models adopted by
Schuler et al, as can be seen in Figure \ref{fmodels}, which 
shows the difference of the thermal structure of the two models for
one of our star in our sample (HE1152-0355). The discrepancy in T reaches 100K.  
Abundances derived from lines originating in the outer
layers of the stellar atmosphere are thus lower in the spherical model
case than in the plane parallel one.  However, such affect is expected
to affect strong lines (in our case mostly CO lines), while its effect
on weak lines such as HF is expected to be very small.
 
As discussed in Section \ref{sec_atm}, for the HF molecular transitions we
use the data in \cite{decin00}, while Schuler et al. use those in
\citet{jorissen92}.  While the $gf$`s for the HF(1-0) R9 line (the
only present in the Schuler et al. spectral range) and dissociation
potential adopted are virtually identical, the excitation potential
for the transition in \citet{decin00} is 0.227\,eV, well below the
0.48 adopted by Schuler et al.  Such difference leads to an offset in
the F abundances measured from the HF(1-0) R9 line.  To asses such
difference, we derived the F abundances adopting the same molecular
transition parameters as Schuler et al., and found that the values
obtained are {\it at most} 0.3\,dex larger using the
\citet{jorissen92} excitation potential than with the \citet{decin00}
one.  Hence, even taking into account the offset due to the difference
in adopted excitation potential, the F abundance and [F/Fe] ratio
measured in HE~1305+0132 are higher than any of the measurements
or upper limits derived in our sample.

It is noteworthy that the Fe abundance in HE~1305+0132 adopted in \citet{schuler07} 
was quite uncertain. In fact, given the limited spectral coverage in the Phoenix
data and the lack of Fe features in that range, 
the authors relied on [Fe/H] measured from intermediate
resolution spectra analysis \citep{beers07}, which reported an
[Fe/H]=$-$2.5$\pm$0.5\,dex.  More recent, high-resolution optical data
suggest that HE~1305+0132 is in fact more metal rich,
with [Fe/H]$=-$1.92\,dex \citep{schuler08}. When taking into account this 
Fe abundance the [F/Fe]
ratio lowers by $\sim$0.6\,dex, to a similar level to those
measured in our sample. Moreover, the F abundance itself, $\log
\epsilon$(F) can also be affected as the adoption of an appropriate,
more metal-rich model atmosphere, whose outer layers are cooler than
those of the metal-poor model atmosphere adopted previously, is
expected to result in a lower F abundance. 

Although the present sample 
shows  that there is a range in [F/Fe] abundances
in CEMP stars, it is not yet clear what the maximum F production is, 
and larger number of
F abundance measurements in CEMP (and EMP) stars is crucial to 
probe the matter.
 
\subsection{Theoretical AGB predictions for F production}\label{sect_compa}
Since \citet{jorissen92},  considerable effort has been 
expended to model its production and explore 
the effect of uncertainties in the adopted parameters.
The main issue to address was that the nucleosynthetic AGB
models failed to account for the high [F/Fe] and [F/O] rations
observed in \citet{jorissen92} in metal-rich intrinsic and extrinsic
C-stars. While the recent results of \citet{abia08} and \citet{abia10} seem to have 
reconciled observations and model predictions at solar and slightly 
subsolar metallicity, large uncertainties in
the $^{19}$F production in AGB stars, especially in the low-metallicity regime,
still exist.
 
\citet{jorissen92} proposed as a main mechanism to produce F in AGB
stars the nucleosynthesis chain
$^{14}$N($\alpha$,$\gamma$)$^{18}$F($\beta^+$)$^{18}$O(p,$\alpha$)$^{15}$N($\alpha$,$\gamma$)$^{19}$F
during the thermal-pulse (TP) in the He intershell.  In this case,
protons for the $^{18}$O(p,$\alpha$)$^{15}$N reaction are provided by
$^{14}$N(n,p)$^{14}$C, and neutrons are provided by 
$^{13}$C($\alpha$,n)$^{16}$O, where $^{13}$C is present in the
H-burning shell ashes at the TP activation. \citet{goriely00} 
proposed that additional F could be made 
during the interpulse phase, where the reaction 
$^{14}$N(n,p)$^{14}$C($\alpha,\gamma$)$^{18}$O(p,$\alpha$)$^{15}$N occurs, 
the neutrons being provided by the $^{13}$C pocket and the protons recycled 
from the $^{14}$N(n,p)$^{14}$C. The production of $^{19}$F is then completed 
in the next  pulse by $\alpha$-capture on $^{15}$N.
\citet{lugaro04} described the
consequence of the reaction rates uncertainties and of the inclusion
on a $^{13}$C pocket on the $^{19}$F yield, showing that their
combined effect can affect it by as much as 40\% in low-mass stars
(M$<3$\,M$_{\odot}$).  \citet{karakas08} instead focused on the
uncertainties on $^{19}$F production stemming from the
$^{18}$F($\alpha$, p)$^{21}$Ne reaction, whose rate was determined
experimentally only in 2006 \citep[see][]{lee06}. Such a reaction was
neglected in the previous work \citep{lugaro04}.  The effect of the
reaction uncertainties on F production is found to increase with
decreasing metallicity, reaching a factor of two for a 2\,M$_{\odot}$
star at [Fe/H]$=-$2.3\,dex.  Another important source of uncertainty is mass
loss.  \citet{stancliffe07} explored the effects of different
prescriptions for mass loss at the beginning of the thermally
pulsing-asymptotic AGB phase on F production, showing that it could
change it by as much as a factor of $\sim$5 \citep[see
also][]{stancliffe08a}.  In particular, \citet{stancliffe08b}, without
any $^{13}$C pocket, using a given mass loss prescription, a different
third dredge-up efficiency from the \citet{lugaro04} and
\citet{karakas08} models, at M=2\,M$_{\odot}$ and [Fe/H]=$-$2.3
predicts an F abundance which is $\sim$10 times lower than than the
\citet{lugaro04}.

All these three family of models
mentioned, however, do not take into account the effects induced by
carbon enhancement on the opacities of the cool external layers of AGB
stars.  When the star becomes C-rich and the number or C ions exceed
that of O ions, the opacity of the external layers is affected
(increased) by the massive presence of C-bearing molecules (mostly
CH, C$_2$, CN and CO). This is expected to causing the envelope to expand and
the star to become larger and cooler \citep[see][]{marigo02}.
\citet{cristallo07} (see also http://www.ugr.es/~cristallo/data\_online.html) 
calculated models of low metallicity, low-mass AGB
stars including the effects of C- and N-enhancements on stellar
opacities.  This has the effect of decreasing the
$^{19}$F production, by roughly a factor of 5 with respect to the
standard models (i.e. obtained using the recommended
reaction rates) from \citet[][]{karakas08}; it is noteworthy that using a solar-scaled composition
Cristallo finds the same results as the standard Karakas model.  It
should be noted that while the in the \citet{lugaro04}, \citet{karakas08} and
\citet{cristallo07} models $^{19}$F production peaks at around
2.2\,M$_{\odot}$, in the \citet{stancliffe08b} it is reached at
1.5\,M$_{\odot}$.

\subsection{Comparison with models: C, N, O and C isotopic ratios} \label{sect_cno}

\begin{figure}
\epsscale{.65}
\plotone{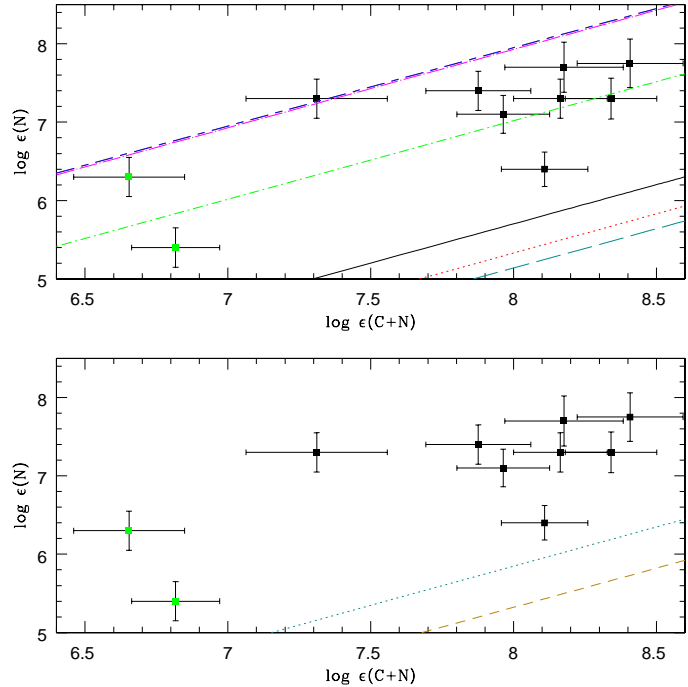}
\caption{Upper panel: N abundance as a function of C+N in 10 CEMP stars.  
Green symbols are CEMP-no stars.
Theoretical modeling results are indicated by lines. Upper panel: plotted models are from \citet{karakas07} 
for different masses. Solid black line 1.25\,M$_{\odot}$, red dotted line 
1.75\,M$_{\odot}$, teal dashed line 2.25\,M$_{\odot}$, green dot-dashed 2.5\,M$_{\odot}$,
magenta dot-dashed line 3.0\,M$_{\odot}$ and 
short-long dashed blue line 4\,M$_{\odot}$.
Lower panel: Model predictions from Cristallo et al. (2009) for a 2\,M$_{\odot}$ star
at a metallicity of -1.3 (dotted teal line) and -2.3 (dashed gold line). 
\label{fcnvsn}}
\end{figure}

\begin{figure}
\epsscale{.65}
\plotone{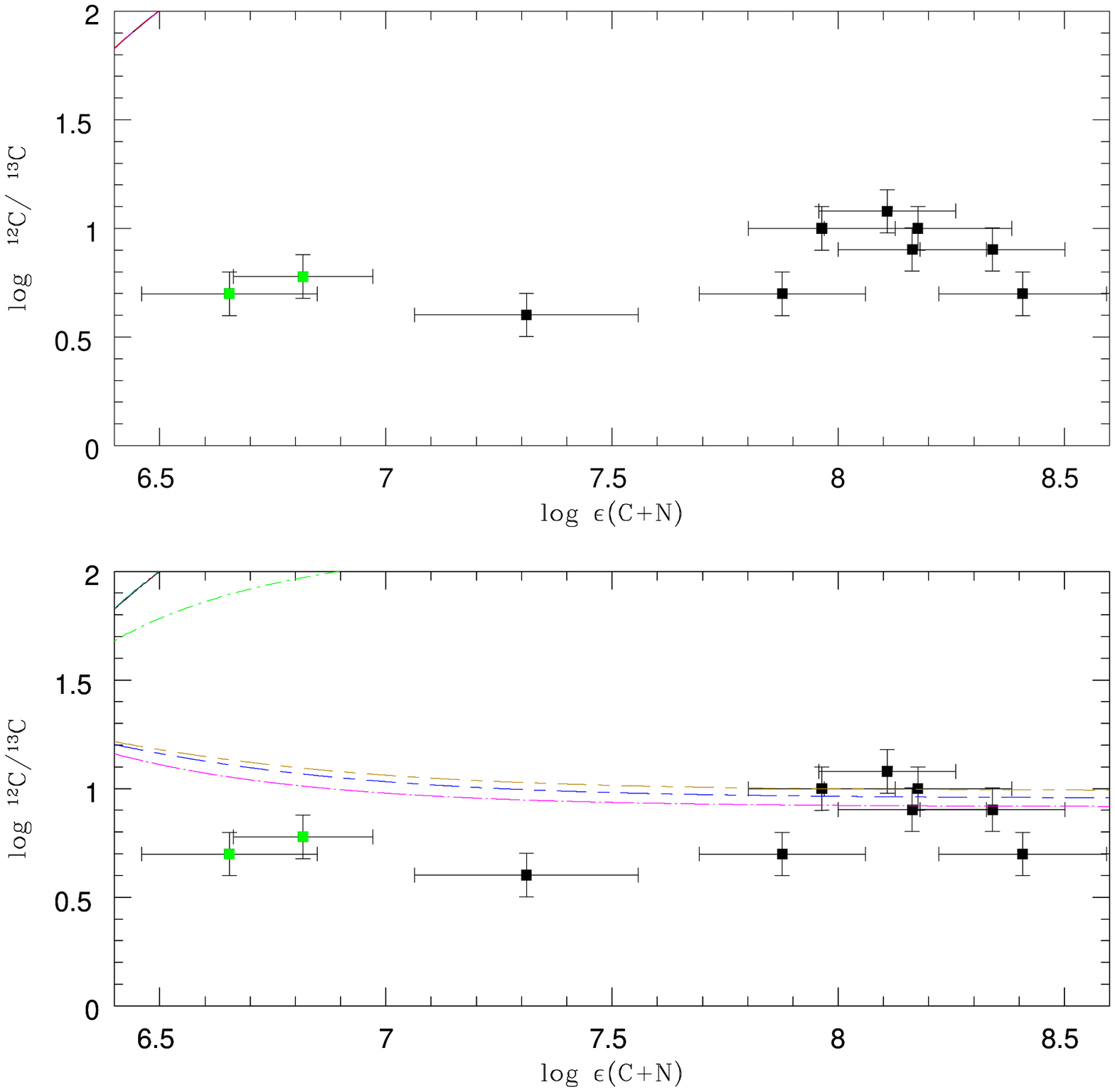}
\caption{Logarithm of C isotopic ratio plotted as a function of C+N. 
Upper panel: theoretical modeling results are indicated by lines. Prediction from \citet{karakas07}, \citet{karakas08}, 
\citet{cristallo07} and \citet{stancliffe08b} are plotted. The models are indistinguishable in the plot as they all predict very similar (high) C isotopic ratios. All models are for M=2\,M$_{\odot}$ and [Fe/H]=$-$2.3.
Symbols in the lower panel are as in Fig. \ref{fcnvsn}, upper panel.
Note that in this case the initial abundance (pre mass transfer) of the observed stars for 
$^{13}$C is taken into account (see text for details).
\label{fcnvsciso}}
\end{figure}

In this section, the results from the previously discussed four different families of models (which provide predictions of mass production of
various elements among which C, N, O, Na and F) are compared to the data. Under the assumption that
the CEMP-s stars observed have been polluted by mass transfer from an
AGB companion and that the amount of C, N, Na, and F in the original (pre
mass-transfer) atmospheres are negligible with respect to that
accreted, the [N/C+N]\footnote{Note that the values of C, N, Na, C isotopic ratios in the following discussion are those listed in Table \ref{tab2} and \ref{tab4}}, [F/C+N] and [Na/C+N] ratios observed in CEMP stars should reflect that
produced by the AGB star companions, independently from the amount of
dilution (ratio between the mass accreted and that of the
envelope of the observed star) undergone by the accreted material.
{\bf Figures \ref{fcnvsn} and \ref{fcnvsciso} show the behaviour of 
N and $^{12}$C/$^{13}$C in the sample stars.  CEMP-s stars are plotted in black and  CEMP-no stars in green.
Given the lack of understanding of the nucleosynthetic
process to which the latter owe their peculiar composition, 
the inclusion in the subsequent plots 
of the two CEMP-no in the comparison to the AGB models is, strictly speaking, not
appropriate.  However, since one of the possible scenario is indeed
analogous to that invoked for CEMP-s stars, we include them in
the plots for completeness (using a different color for easy differentiation).
In Fig. \ref{fcnvsn}, for each individual model different degrees of dilution 
correspond to different positions along its correspondent line.

The choice of the sum of the two species C+N rather than a
single one at a time is because it is a much more solid
prediction in stellar evolution models.  It is well known
that the standard AGB models fail to reproduce the N abundances
observed in C-rich stars \citep[see e.g.][]{jorissen92}.
We find similar results in our sample. The common explanation of such apparent disagreement
is that extramixing processes take place at the bottom of the AGB envelope,
where part of the envelope material undergoes the CN cycle, turning
carbon into nitrogen (see the discussion at the end of this section).

 As can be noticed from Fig. \ref{fcnvsn} 
the low mass models predict a N abundance which is lower than the measured ones,
which are closer to the predictions of more massive AGB stars (2.5-4\,M$_{\odot}$) models,  where HBB takes place. 

A similar conclusion can be drawn from Fig. \ref{fcnvsciso}, which plots 
$^{12}$C/$^{13}$C with respect to C+N. It is worth noting that, unlike the previous case with N and those of F and 
Na that will be discussed later, the initial composition of the star (before mass transfer) is not 
negligible when  plotting the model prediction.
In fact, while essentially all C (as $^{12}$C) comes from mass transfer, according to the low-mass AGB models (which do not include extramixing processes) the $^{12}$C/$^{13}$C ratios are very high, thus according to these models, very little $^{13}$C is transferred with respect to $^{12}$C (the predicted isotopic ratio can be 
as high as 10$^4$, hence 10$^4$ $^{12}$C nuclei are transferred for a single $^{13}$C nucleus). The amount of $^{13}$C
in the atmosphere of the CEMP star before it becomes CEMP counts for small amounts of mass transfer.
Fig. \ref{fcnvsciso} shows the predictions for the same models shown in Fig. \ref{fcnvsn} under the assumption 
that the original (pre mass-transfer) C abundance and C isotopic ratios in the observed stars 
were [C/Fe]=-0.4 and 20, which is a typical value measured for metal-poor giants (see e.g. \citet{gratton00}).
The models in the plot take into account different degrees of dilution, corresponding to different positions 
along the lines.
The C isotopic ratios measured in our sample are  all quite low,
some almost at the equilibrium value, much lower than predicted by low-mass
shell-nucleosynthesis models \citep[$\sim$10,000 see e.g.][]{cristallo07}, while the 
ratios for more massive stars ($\geq$2.5\,M$_{\odot}$) are closer to the measured ones.
However, a star polluted by a star with a mass $\geq$2.5\,M$_{\odot}$, where HBB operates should be characterized by [N/C]$>$0, which is only found in two of the stars in our sample, making this 
possibility quite unlikely (see also the discussion in the next section).

During their evolution, the stars in our sample undergo several mixing episodes 
(i.e. first dredge-up, extramixing corresponding to the RGB bump) from the
main sequence to the evolutionary state we observe them in today.
This mixing episodes are known to affect the C isotopic ratio in C normal stars. 
It is reasonable to assume that they also might affect the accreted C, altering the 
C isotopic ratio of the accreted material.

This effect is however likely not important.
In fact, similar values for C isotopic ratios are found in unevolved 
CEMP-s stars (see e.g. \citealt{lucatello03,masseron09a,aoki02}), which have not undergone any of such mixing episodes and have thus maintained the original $^{12}$C/$^{13}$C in transferred material. \newline
 Moreover,  the extent of the effect of the mixing episodes on the accreted C isotopic ratio should likely depend on the time in its evolution when the C-rich material is transferred. Since it is sensible to assume that the CEMP-s stars in 
our sample underwent mass transfer at different times in their evolution, the fact that all the measured isotopic ratios are all quite low, argues that the effect of the mixing episodes on such quantity should be marginal.
%
Therefore, it is reasonable to conclude that the measured C isotopic ratios indeed 
reflect that in the polluting material.

To account for such a high N abundance and low C isotopic ratios in low-mass AGB stars, we confirm that  
some degree of of extra-mixing with a partial activation of proton capture nucleosynthesis,
the so-called Cool Bottom Processing (CBP), needs to
be invoked \citep[][]{nollett03}.
However, in most published works, such process is treated as a 
parameter and thus the amount of
CN processing is highly uncertain, making the individual C and N
abundance predictions much more sensitive to modeling details than
their sum. This process is
known to occur during the first red giant phase in stars with
M$\leq$2.5\,M$_{\odot}$  \citep{charbonnel95}
and was suggested to occur between dredge 
up in AGB stars \citep[][]{nollett03}. It consists of
circulation of material from the base of the convective envelope into
the thin radiative region located on top of the H-burning shell, where
the material is processed by proton captures and then brought back to
the envelope, producing the signature of CN processing at the stellar
surface. 
The low $^{12}$C/$^{13}$C ratios, high N (but with [C/N]$\geq$0)
observed in CEMP-s stars indicates that CBP also takes place in low
metallicity AGB stars \citep[see also][]{aoki02,masseron09a}. 
Unfortunately, the physical mechanism responsible for such process is still not clear.
Several scenarios have been proposed, e.g., the mixing process could be driven by a
molecular weight inversion created by the $^3$He($^3$He,2p)$^4$He 
reaction \citep{eggleton08}, or by 
magnetic buoyancy at the bottom of the envelope (e.g., 
Palmerini, Nollett, \& Busso 2008).

While we obtained measurements for O for all the stars in our sample, 
a comparison with the model predictions would not be meaningful. All the models discussed in the text start with solar-scaled composition. While this is not important for the other elements we discuss as their 
production in the AGB nucleosynthesis is much larger than the initial content, this is not true for O, which is in fact 
typically produced at the [O/Fe]$\sim$0.5 level, a value comparable to that measured in metal-poor stars,
 making the adoption of a correct initial O for the models 
important. The typical predicted values for O are in fact lower than the measured ones; however it is presently not possible to determine whether this is due to the incorrect assumption on the model star initial composition or issues in the understanding of O nucleosynthesis in low-metallicity AGB stars (see for instance
the impact of overshooting on O abundance, e.g., Herwig 2005 ARA\&A).
}
\subsection{Comparison with observations: F}\label{sect_comp_f}

F was measured only in two CEMP stars in our
sample, both s-process rich, while for the remaining eight objects
(including the two CEMP-no stars) only upper limits could be placed.
Therefore, as far as $^{19}$F is concerned, no meaningful comparison is possible
between CEMP-s and CEMP-no stars, and not much information
on the formation scenario of CEMP-no stars can be derived from the
present data.  

Fig. \ref{fig4} shows the derived F abundances with respect to the
sum of the C and N abundances in the sample stars.  

Results from four different families of models are plotted for
comparison to the data; they provide yields for of
various elements among which are C, N, Na, and F. 
In Fig. \ref{fig4}, for each individual model different degrees of dilution 
correspond to different positions along its correspondent line.

In the upper panel the model plotted are all computed for a mass of 2M$_{\odot}$ and a metallicity of [Fe/H]=$-$2.3.  The dotted line indicates the prediction
from \citet{karakas07}, the one used in \citet{lugaro08} to
account for the F content reported in \citet{schuler08}.  The solid
line describes the model from \citet{karakas08}, which includes 
a $^{13}$C pocket of 0.002M$_{\odot}$ at the top layer of the He- and C-rich inter-shell region and adopts the upper limit for the rate of the $^{18}$F($\alpha$, p)$^{21}$Ne reaction (in an
attemp to reproduce the high F abundances in the Jorissen
sample).
Among the models available in the literature, this is the one
that for the mass and metallicity  under consideration, produces the largest
amount of F with respect to C and N.  The
dot-dashed line indicates the prediction of the model from \citet{cristallo07}., 
which accounts for the appropriate C-bearing molecules opacities in
the stellar evolution model, unlike the two other models, which adopt solar
scaled composition.  Finally, the dotted line shows the prediction from
the \citet{stancliffe08b} model, which uses solar-scaled opacities.
 
As shown in the upper panel of Fig. \ref{fig4}, the
two F measurements can be accounted for by the 
\citet{karakas07} or the Cristallo et al. (2009) 2 M$_{\odot}$ 
star models.
On the other hand, the F predicted from the 
Karakas et al. (2008) 2 M$_{\odot}$ star is higher.
All the upper limits excepting one derived for F 
in our sample fall lower than the two measured abundances.
Two of the upper limits are in agreement with the predictions of 
Stancliffe \& Glebbeek (2008) in the figure.
We note that a significant fraction of the upper limits 
are located under the predicted lines.

As mentioned before, for the Lugaro et al., Cristallo et al. and 
Karakas et al., F production peaks at $\sim$ 2.2\,M$_{\odot}$
\citep[see][]{lugaro08}. Quantitative predictions are provided only
for 2.0\,M$_{\odot}$ AGB models in \citet{cristallo07} and 
\citet{karakas08} papers, so no
comparison with different mass models could be performed.  On the
other hand, \citet{stancliffe08a} prediction is lower than the measurements,
possibly accounting for two of the upper limits (one CEMP-s and one CEMP-no star).
However, it still cannot account for most of the upper limits derived
for CEMP-s stars. In the \citet{stancliffe08a} models, unlike the others, F production peaks
at $\sim$1.5\,M$_{\odot}$.  Hence, we cannot invoke lower mass models 
to reconcile the discrepancy in this case.

\citet{karakas07} provide F production predictions for several AGB masses.
In the lower panel of Fig. \ref{fig4} the predictions for the \citet{karakas07} models 
are plotted together with the data. 
Models are for masses 1.25, 1.75, 2.0, 2.5, 3.0, 3.5 and 4\,M$_{\odot}$ at a metallicity 
of [Fe/H]$=-2.3$.
The models at 1.25, 1.75 and 2.5\,M$_{\odot}$ seem to reproduce satisfactorily the 
two F measurements in our sample, even though it should be kept in mind that, as mentioned,
their metallicity is higher than that of the models ([Fe/H]=$-1.3$ and $-2.0$).
However, as discussed later in this section, the effect of metallicity on [F/C+N] is likely very small.
On the other hand, when considering the measurements obtained for the whole sample, 
it can be noticed that predictions for F to C+N for 
masses from 1.25 to 2.5\,M$_{\odot}$ are higher 
than four out of the eight CEMP-s stars.
While the predictions for models for 3, 3.5 and 4\,M$_{\odot}$ can account 
for the upper limits F to C+N ratios in those four CEMP-s, 
it is important to keep in mind that for [Fe/H] $\lesssim -$2 all AGB stars 
with intial mass higher than M$\sim$2.5\,M$_{\odot}$ will be affected by HBB,
which burns C producing N (e.g., Karakas \& Lattanzio 2007).
 A star polluted by such an object, would not be a C-rich, but rather N-rich,
which is not the case of our sample (three out of those five
CEMP-s have [C/N]$>$0).
Moreover, stars of this mass range are not very efficient in producing
s-process elements (which are enhanced in eight stars in our sample,
we provide Ba abundances in Table \ref{tab2} for completeness).
Finally, a simple back-of-the-envelope calculation  adopting a Salpeter IMF 
and assuming that, in the considered mass range, the likelihood of 
having a companion is not affected by mass, 
indicates that the probability of observing, among a sample of eight
CEMP-s stars with a low-mass companion (1.25$<$M$<$4\,M$_{\odot}$),
 four object that had a companion in the 2.5 to 4\,M$_{\odot}$ mass range,
is very small (10$^{-2}$). 
For these reasons, the possibility that the abundance ratios observed are due
to stars more massive than 2.5\,M$_{\odot}$ seems highly unlikely. 
{\bf We note that for one of the CEMP-no stars, CS\,30314-67, the derived upper limit to F 
abundance is also quite low and, similarly to most CEMP-s stars in our sample,
 is not reproduced by the low-mass, low-metallicity AGB models. This 
could be in principle used as an argument against the scenario which attributes 
their peculiar composition to pollution from low-mass AGB. However, 
given the issues that the models have in reproducing the observations 
as far as F is concerned in CEMP-s stars (which are known to arise from AGB mass transfer),
the AGB pollution scenario cannot be ruled out.}

Notice that for each model discussed we consider the last thermal pulse. 
According to e.g. Izzard \& Tout (2003)
AGB evolution can be truncated by binary interaction, 
with the AGB star experiencing fewer thermal pulses
with respect to predictions of an isolated object.
To illustrate the evolution of the abundance
ratios through different thermal pulses we plot $\log \epsilon$(F) and
$\log \epsilon$(C+N) from 2\,M$_{\odot}$ (Cristallo
et al. 2009), overplotting our measurements for comparison (see upper panel of 
Fig. \ref{fig4a}). With the exception of one limit, 
in principle most of the upper limits could be explained by 
predictions of this particular model at different thermal pulses.
 In the lower panel, on the other hand, 
we show predictions from Cristallo et al. (2009) for a 2\,M$_{\odot}$
with two different metallicities, [Fe/H]=$-$1.3 and $-$2.3. The figure 
shows a change of an order of magnitude in metalicity
has only a small effect on the ratio [F/C+N]. For this reason, even though 
our sample star cover a large range in metallicity, we compare our measurement 
to the predictions at a single metallicity, the lowest one available. 
We invite the reader to keep in mind that this is an extrapolation 
to lower metallicity of the behaviour of the model and as such 
quite uncertain.

Beside the sources of uncertainties discussed in the previous
paragraphs, as mentioned before, currently
AGB nucleosynthesis calculations do not include CBP. 
Indeed, at typical AGB envelope conditions F can be depleted significantly 
by extra-mixing processes, depending on their efficiency 
(Palmerini, Nollett, \& Busso 2008), via the $^{19}$F(p,$\alpha$) reaction.
Therefore, the inclusion of an appropriate treatment for CBP in 
AGB modeling could possibly decrease the F production, bringing the predictions 
closer to the observations.
It is noteworthy that  partial F depletion has been proposed to be active at the tip of the previous Red Giant Branch (RGB) phase (Denissenkov et al. 2006). It must be 
kept in mind, however, that such a mechanism has been invoked to explain 
the decrease of F abundance in stars at the tip of the RGB in the 
globular cluster M\,4, which, as likely all globular clusters, 
show abundance variations which are not observed in field stars \citep[see e.g.][]{carretta09}. It is also important to note that 
one of the two stars in our sample for which we can actually measure F,
HE\,1152-0355, is quite close to the RGB tip.

\subsection{Comparison with observations: Na} \label{sect_comp_na}
Models for low-mass AGB 
stars predict quite successfully the observed Na abundances.
Na is produced by AGB stars in different ways.
First, during the TP neutrons produced via
the $^{22}$Ne($\alpha$,n)$^{25}$Mg may be captured by the 
$^{22}$Ne itself, producing $^{23}$Na via the reaction chain
$^{22}$Ne(n,$\gamma$)$^{23}$Ne($\beta^-$)$^{23}$Na.
Secondly, sodium may be strongly produced also by $^{22}$Ne(p,$\gamma$)$^{23}$Na.
Indeed, after the TDU, at similar mass coordinate of the $^{14}$N-pocket a
$^{23}$Na-pocket is formed in radiative conditions, where $^{23}$Na 
is given by the equilibrium between the production channel
$^{22}$Ne(p,$\gamma$)$^{23}$Na and the depletion channels
$^{23}$Na(p,$\gamma$)$^{24}$Mg and $^{23}$Na(p,$\alpha$)$^{20}$Ne
(see Cristallo et al. 2006 for details, and Bisterzo et al. 2006).
Finally, a marginal contribution is given by neutron capture on 
$^{22}$Ne in the $^{13}$C-pocket.
Despite these different processes, Na production in AGB stars is less affected by
nuclear uncertainties compared to F. 
For these reasons, AGB predictions for Na are expected to be more robust than F. 
 Fig. \ref{fig10} shows Na and C+N in the sample stars together with the predictions 
from the same models as in Fig. \ref{fig4}.
The upper panel shows two of the models represented in the correspondent panel 
of Fig. \ref{fig4} because those are the only ones that provide Na abundance predictions.
It is noteworthy how the discrepancy in Na between these two models ($\sim$0.05\,dex at a 
given $\log \epsilon$ (C+N) is much smaller 
than that in F between the same models ($\sim$0.5\,dex), supporting the idea  
that Na production is generally better understood and/or less sensitive to the 
recipes and parameters adopted in different families of models.
As can be noticed from in Fig. \ref{fig10b}, 
the measured Na abundances are consistent with pollution arising from a 1.25 to 
1.75\,M$_{\odot}$ AGB star. Six out of seven sample 
CEMP-s for which Na was measured and both CEMP-no stars 
are consistent with having a companion in that mass range.

\section{Conclusions}\label{conclusions}
We have obtained IR high-resolution observations in a sample of
ten CEMP stars, eight CEMP-s and two CEMP-no. The aim was to measure Na abundances, 
 $^{19}$F contents from HF, O from CO and N from CN lines in the $K$
band.  

A comparison of predictions from different families 
of low-metallicity, low-mass (1.25 to 1.75 M$_{\odot}$ range) AGB nucleosynthesis models 
show that they reproduce well the observed Na abundances in the CEMP-s stars 
in our sample. This result strongly argues in 
favor of a polluter of low-mass for these objects, in agreement with the 
currently most accepted scenario 
for their formation.

F could only be measured in two CEMP-s stars, while for the
remaining eight objects only upper limits could be derived. However, these
data are sufficient to show that a range of [F/C+N] values
are produced in low-metallicity AGB stars, in accord with predictions
of the mass- and metallicity-dependence of F production in AGB models.

\begin{figure}
\epsscale{.65}
\plotone{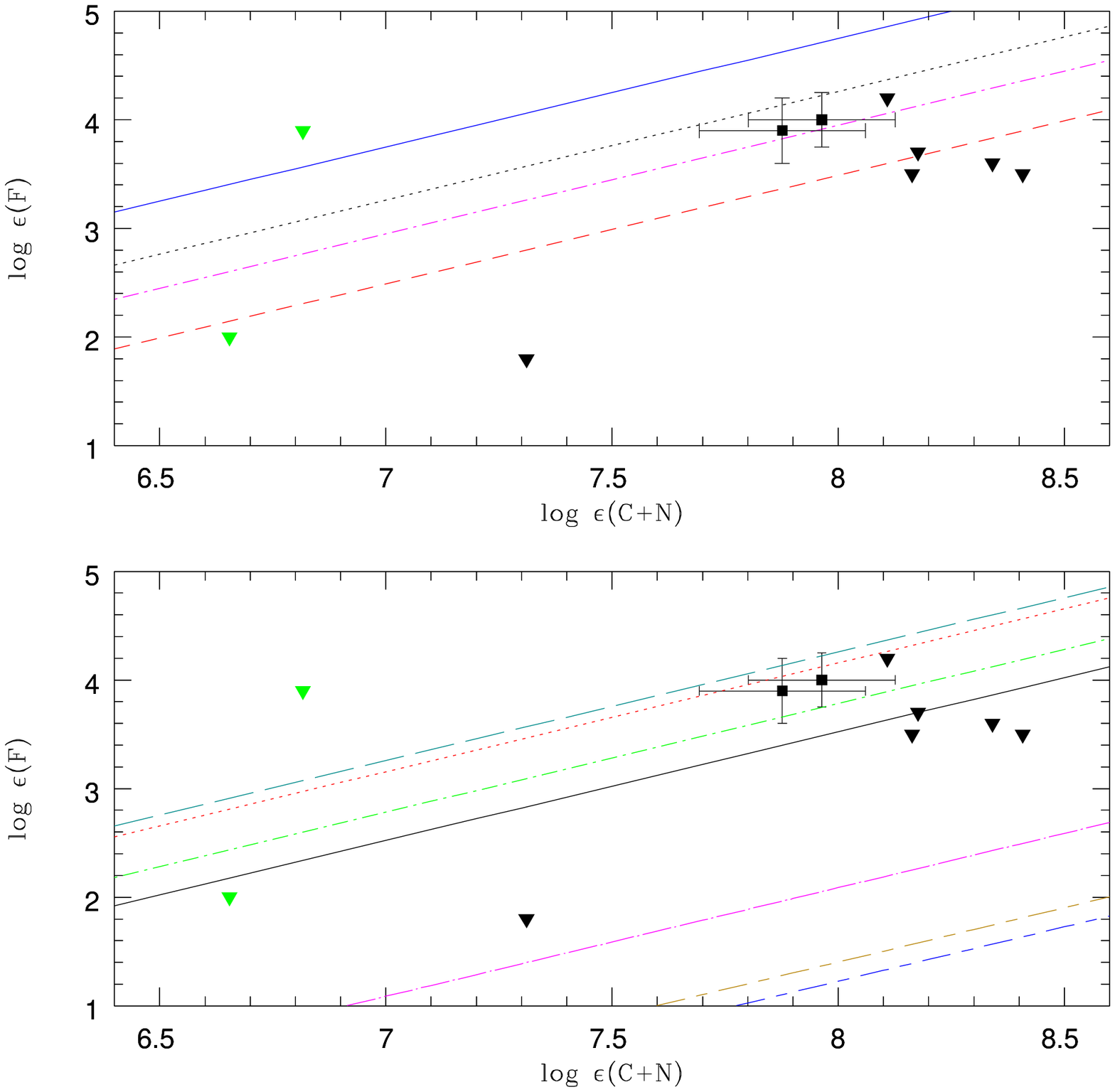}
\caption{
Upper panel: F abundance as a function of C+N in 10 CEMP stars. The inverted triangles indicate upper limits, 
green symbols are CEMP-no stars.  Theoretical modeling results are indicated by lines. Dotted
line indicates the prediction from \citet{karakas07}, the solid line is
from \citet{karakas08}, the dot-dashed line is from
\citet{cristallo07} and the dashed line is from
\citet{stancliffe08b}. All models are for M=2\,M$_{\odot}$ and [Fe/H]=$-$2.3.
Lower panel: as in upper panel, but plotted models are from \citet{karakas07} 
for different masses. Solid black line 1.25\,M$_{\odot}$, red dotted line 
1.75\,M$_{\odot}$, teal dashed line 2.25\,M$_{\odot}$, green dot-dashed 2.5\,M$_{\odot}$,
magenta dot-dashed line 3.0\,M$_{\odot}$, short-long dashed yellow line 3.5\,M$_{\odot}$  and 
short-long dashed blue line 4\,M$_{\odot}$.
\label{fig4}}
\end{figure}

\begin{figure}
\epsscale{.65}
\plotone{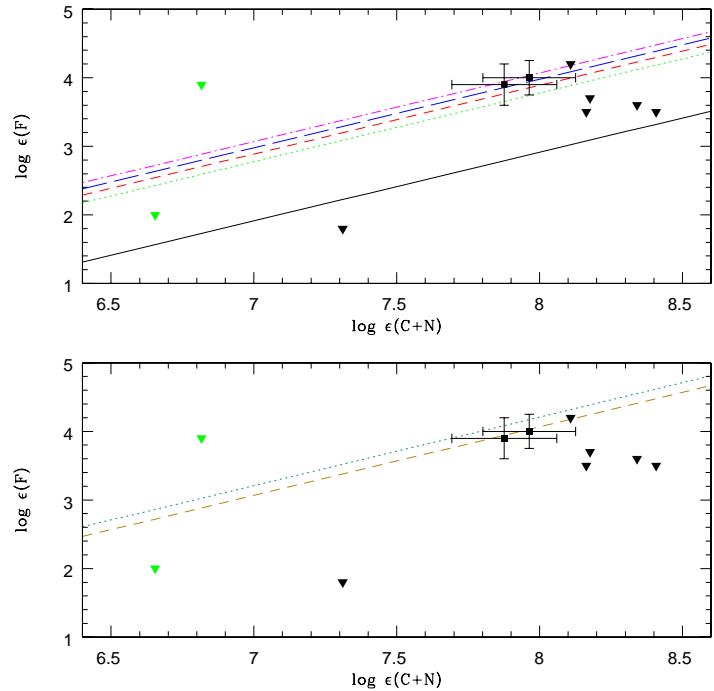}
\caption{Symbols for measurements in the sample are as in Fig. \ref{fig4}. 
Upper panel: $\log \epsilon$(F) and $\log \epsilon$(C+N) for a 2\,M$_{\odot}$ star at 
[Fe/H]=-2.1 (Cristallo et al. 2009) for thermal pulses 1 (black solid line), 3 (dotted green line), 5
(short-dashed red line), 10 (long-dashed blue line) and 15 (last pulse, dot-dashed magenta line). 
Lower panel: Model predictions from Cristallo et al. (2009) for a 2\,M$_{\odot}$ star
at a metallicity of -1.3 (dotted teal line) and -2.3 (dashed gold line). 
\label{fig4a}}
\end{figure}

\begin{figure}
\epsscale{.65}
\plotone{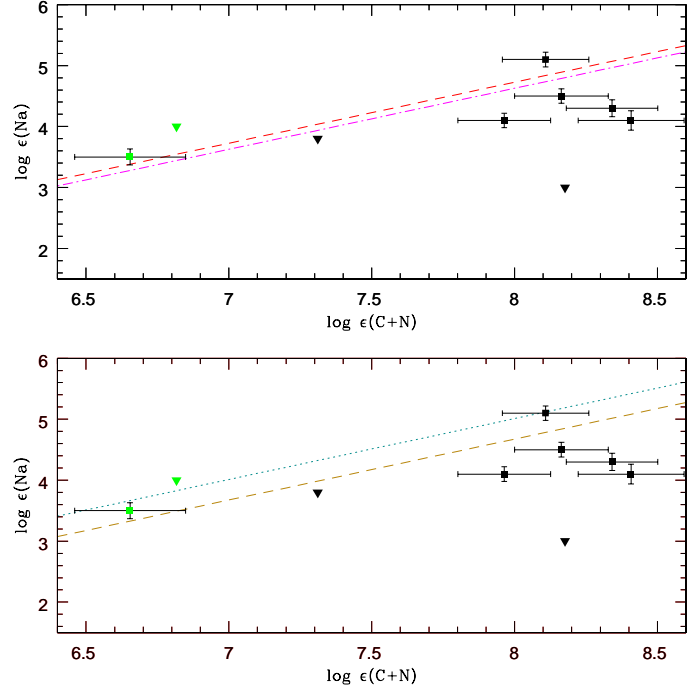}
\caption{Na plotted as a function of C+N. Symbols as in Fig. \ref{fig4}.
Note that only two of the four models plotted in the upper panel of Figure \ref{fig4} 
are plotted here as they are the ones reporting predictions for Na production.
The lower panel shows Na production prediction in \citet{cristallo07} at [Fe/H]=$-1.3$ and $-2.3$.
\label{fig10}}
\end{figure}

\begin{figure}
\epsscale{.65}
\plotone{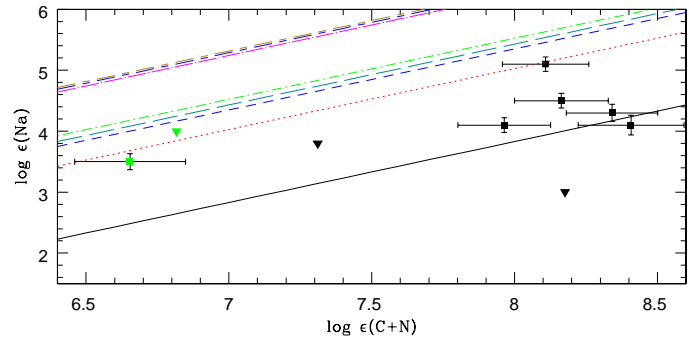}
\caption{Na plotted as a function of C+N compared with different 
model predictions. Symbols as in Fig. \ref{fig4}.
\label{fig10b}}
\end{figure}

The only two F measurements obtained (the CEMP-s stars 
HE\,1152-0355 and HD\,5223 respectively at [Fe/H]=$-$1.27 and 
$-$2.06) are accounted for, within the errors, low-mass, low-metallicity AGB models.  
On the other hand, most of the derived upper limits for 
 F abundance in CEMP-s are not satisfactorily accounted for by nucleosynthetic
computations.  In fact, the comparison with four of the most recent
models for low-mass (2.0\,M$_{\odot}$), low-metallicity AGB
nucleosynthetic models shows that there are large differences in the predictions 
between different families of models, which cannot reproduce several of the 
upper limits, not providing any [F/C+N] ratios predictions low enough 
to account for the values measured in several of the sample stars.
A comparison with \citet{karakas07} models for different masses indicates 
that only objects more massive (2.5\,M$_{\odot}<$M$<$4\,M$_{\odot}$) 
that those generally considered as responsible for CEMP-s peculiar abundances 
(M$\simeq$1.5\,M$_{\odot}$)
produce the [F/C+N] ratios observed in about two thirds of our CEMP-s sample. 
This possibility is not only unlikely because of simple IMF considerations, 
but it is also challenged by the fact that at low metallicity, HBB is active 
in stars in this mass range, producing N and depleting C (bringing the ratio 
to [N/C]$\sim$1 \citep[see e.g.][]{johnson07}) and preventing the star from becoming 
a CEMP-s polluter. {\bf As shown in Fig \ref{fig4}, the \citet{karakas07} 1.25\,M$_{\odot}$ model 
is the low-mass AGB ones that comes closest to explaining the upper limits. 
Given the discussed spread present in F production in different models,  
it is indeed possible that it affects also predictions for other masses, including 1.25\,M$_{\odot}$ , 
hence accounting for the derived upper limits. 
It is however impossible, at present time, to systematically test this differences, as only \citet{karakas07} provide 
F production calculations for several AGB masses.}

We discussed two possible solutions that could explain the lowest F upper limit
range. One is that the standard evolution of an AGB star may be truncated by binary interaction, and 
a lower F abundance can be obtained in early TP pulses (even though this would lead to lower 
Ba production, which could make the accounting for observed Ba abundances possibly more challenging).
The other is the action of  CBP, which may reduce the F in the AGB envelope if such extra-mixing processes
exposes material at temperatures high enough to activate $^{19}$F(p,$\alpha$).

The conclusion that can be drawn from the comparison between the data 
and the current models is that F production in low metallicity AGB stars 
is probably not as high as expected on the basis of the current models.
It is important to keep in mind that, as discussed in the introduction, 
at solar metallicity model predictions are in agreement with the observations, 
therefore such problem seem to be peculiar to the low metallicity regime.
AGB nucleosynthesis, especially at low metallicity, 
is still not fully understood and a large number of uncertainties affect in particular 
F nucleosynthesis.
Improvements in nuclear reaction rate (such as $^{14}$C($\alpha$,$\gamma$)$^{18}$O
and $^{18}$F($\alpha$,p)$^{21}$Ne) accuracy are needed, 
and a better grasp of the CBP mechanism is highly desirable.
Only one of the available models, by \citet{cristallo07} 
takes into account the appropriate C, N and O enhancements in the computation of 
the stellar opacities; also, the prescription for mass loss is still rather uncertain. 
Moreover, currently no model takes into account the effect on the AGB evolution of 
the presence of a (close) companion, which could stimulate extra mixing and/or 
truncate the AGB phase (see e.g. Izzard \& Tout 2003), affecting the nucleosynthesis.
The inclusion of these ingredients into a new generation of AGB models might 
in the future be able to account for the present observations.
%

From an observational point of view, pursuing a larger number of 
F abundance measurements in CEMP stars is extremely important. 
In particular, K-band high-resolution spectral observations of 
cool (T$_{eff}<$4200\,K) CEMP stars would allow actual measurements 
(rather than upper limit
placements) of F even for low F abundances ($\log \epsilon$(F)$\sim$2).
 These measurements would provide more stringent constraints that are urgently needed 
for a future generation of AGB models and to effectively probe F chemical evolution.

\acknowledgments
We thank the anonymous referee for helping improving the paper.
This research was supported by the DFG cluster of excellence 
"Origin and Structure of the Universe"(www.universe-cluster.de).
J. A. J. and T. M. acknowledge support from NSF grant AST-0707948.
MP acknowledges support by the Joint Institute for Nuclear
Astrophysics (JINA), NSF grant PHY0822648,  and from the National Research
Council of Canada.

\clearpage




\clearpage
\newpage
\begin{deluxetable}{lrrrrrr}
\tabletypesize{\scriptsize}
\tablecaption{Observational log\label{tab1}}
\tablewidth{0pt}
\tablehead{
\colhead{Star} & \colhead{RA} & \colhead{Dec} & \colhead{V} & \colhead{K} & \colhead{Obs} & \colhead{Exp.}\\
\colhead{ID}   &\colhead{} & \colhead{} & \colhead{} & \colhead{} & \colhead{Date} & \colhead{Time (s)}
}
\startdata
CS\,22942-19  &00 57 18.136&$-$25 26 09.69&12.69 & 10.645 & 2007-10-21& 1200 \\
CS\,29497-34  &00 41 39.818&$-$26 18 54.48&12.22 &  9.996 & 2007-10-26& 900  \\
CS\,30314-67  &20 52 50.986&$-$34 19 40.52&11.85 &  9.028 & 2007-10-23& 900  \\
CS\,22948-27  &21 37 45.774&$-$39 27 22.31&12.660& 10.427 & 2007-10-27& 1200 \\
CS\,29502-92  &22 22 35.995&$-$01 38 27.53&11.87 &  9.601 & 2007-10-24& 1200 \\
CS\,30322-23  &21 30 32.184&$-$46 16 24.72&12.21 &  9.313 & 2007-10-23& 900  \\
HE\,1152-0355 &11 55 06.057&$-$04 12 24.71&11.43 &  8.429 & 2008-02-26& 720  \\
HE\,1305+0007 &13 08 03.849&$-$00 08 47.40&12.223&  9.600 & 2008-02-27& 1200 \\
HD\,187861    &19 56 26.945&$-$65 22 08.01&9.19 &  6.699 & 2007-10-21& 180  \\
HD\,5223      &00 54 13.610&$+$24 04 01.51&8.47 &  5.673 & 2007-10-21& 180  \\
HD\,122563    &14 02 31.846&$+$09 41 09.94&6.20 &  3.731 & 2008-02-28&  45  
 \enddata
\end{deluxetable}

\begin{deluxetable}{lrrrrr}
\tabletypesize{\scriptsize}
\tablecaption{Adopted atmospheric parameters, C and Ba abundances\label{tab2}}
\tablewidth{0pt}
\tablehead{
\colhead{Star} &\colhead{T$_{\rm eff}$}&\colhead{$\log g$}&\colhead{[Fe/H]}& \colhead{[C/Fe]}&\colhead{[Ba/Fe]}\\
\colhead{ID} &\colhead{(K)}&\colhead{}&\colhead{(dex)} & \colhead{(dex)} & \colhead{(dex)} 
}
\startdata
CS\,22942-19   & 5100$\pm100^a$ &  2.5$\pm0.1^a$ & -2.5$\pm0.15^a$ &2.2$\pm0.10^a$      & 1.76$\pm0.10^b$    \\
CS\,29497-34   & 4800$^{c}    $ &  1.8$^{c}    $ & -2.9$^{c}     $ &2.6$^{c}           $& 2.03$^c$           \\
CS\,30314-67   & 4400$\pm100^b$ &  0.7$\pm0.3^b$ &-2.85$\pm0.18^b$ &0.8$\pm0.18^{b,i}$  &-0.57$\pm0.14^b$    \\
CS\,22948-27   & 4800$^c      $ &  1.8$^c      $ & -2.5$^c       $ &2.4$^c           $  & 2.26$^c$           \\
CS\,29502-92   & 4890$\pm100^d$ &  1.7$\pm0.2^d$ & -3.18$\pm0.09^d$&0.96$\pm0.22^{d,i}$ &-1.26$\pm0.09^d$    \\
CS\,30322-23   & 4100$\pm100^e$ & -0.3$\pm0.3^e$ & -3.39$\pm0.18^e$ &0.6$^e$            & 0.54$\pm0.10^e$    \\
HE\,1152-0355  & 4000$\pm 90^f$ &  1.0$\pm0.3^f$ & -1.27$\pm0.27^f$ &0.58$\pm0.2 ^f$    & 1.58$^f$           \\
HE\,1305+0007  & 4560$\pm 90^g$ &  1.0$\pm0.3^g$ & -2.5$^g$       &2.4$\pm0.35^g $      & 2.9$\pm0.5^g$      \\
HD\,187861     & 4600$\pm150^a$ &  1.7$\pm0.2^a$ & -2.36$\pm0.21^a$ &2.02$\pm0.10^a$      & 1.39  $\pm0.10^a$\\
HD\,5223       & 4500$\pm 90^f$ &  1.0$\pm0.3^f$ & -2.06$\pm0.13^f$&1.57$^f$            & 1.82$^f$           \\
HD\,122563     & 4600$\pm100^h$ &  1.1$\pm0.3^h$ & -2.82$\pm0.10^h$&-0.10$\pm0.05^{d,i}$ &-1.28$\pm$0.23 $^b$
 \enddata
 \tablecomments{\scriptsize{a} \citet{masseron09}, \scriptsize{b} \citet{aoki02},
\scriptsize{c} \citet{barbuy05}, \scriptsize{d} \citet{lai07}, \scriptsize{e} \citet{masseron06},
 \scriptsize{f} \citet{goswami06}, \scriptsize{g} \citet{beers07}, \scriptsize{h} \citet{spite05},
\scriptsize{i} C abundance has been adjusted (see text for details)
}
\end{deluxetable}



\begin{deluxetable}{lcccccccr}
\tabletypesize{\scriptsize}
\tablecaption{Sensitivity of the derived abundances to the atmospheric parameters for HE\,1152-0355 and HD\,5223\label{tab3}}
\tablewidth{0pt}
\tablehead{
\colhead{}&\multicolumn{4}{c}{HE\,1152-0355}&\multicolumn{4}{c}{HD\,5223}\\
\colhead{Element}&\colhead{T$_{eff}$}&\colhead{$\log g$}&\colhead{[Fe/H]}&\colhead{[C/Fe]}&\colhead{T$_{eff}$}&\colhead{$\log g$}&\colhead{[Fe/H]}&\colhead{[C/Fe]}\\
\colhead {}      &\colhead{$+200$\,K}&\colhead{$+0.5$\,dex}&\colhead{$+0.3$\,dex}&\colhead{$0.3$\,dex}&\colhead{$+200$\,K}&\colhead{$+0.5$\,dex}&\colhead{$+0.3$\,dex}&\colhead{$0.3$\,dex}}
\startdata
O & 0.4 & 0.1 & 0.1 & -0.1        & 0.5 & 0.1 & 0.1 & -0.1\\ 
N & 0.4 & 0.1 & 0.1 & -0.2        & 0.4 & 0.1 & 0.1 & -0.2\\
F & 0.5 & 0.0 & 0.2 &  0.1        & 0.5 & 0.0 & 0.2 &  0.1\\
Na&\nodata&\nodata&\nodata&\nodata& 0.2 & 0.1 & 0.1 &  0.1\\
\enddata
\end{deluxetable}

\begin{deluxetable}{lccccccr}
\tabletypesize{\scriptsize}
\tablecaption{N,O, Na and F abundance measurements for the program stars. C isotopic ratio is also listed\label{tab4}}
\tablewidth{0pt}
\tablehead{
\colhead{Star} & \colhead{$^{12}$C/$^{13}$C}&\colhead{$\log\epsilon$(N)}&\colhead{[N/Fe]}&\colhead{$\sigma(\log \epsilon$(N))}&
\colhead{$\log\epsilon$(O)}&\colhead{[O/Fe]}&\colhead{$\sigma(\log \epsilon$(O))}}
\startdata
CS\,22942-19  &12&  6.4 & 1.1& 0.22& 7.2  & 1.0& 0.21\\
CS\,29497-34  & 8&  7.3 & 2.4& 0.25& 7.4  & 1.6& 0.22\\
CS\,30314-67  & 5&  6.3 & 1.4& 0.25& 6.4  & 0.6& 0.23\\
CS\,22948-27  & 8&  7.3 & 2.0& 0.26& 7.3  & 1.1& 0.23\\
CS\,29502-92  & 6&  5.4 & 0.8& 0.25& 6.8  & 1.2& 0.22\\
CS\,30322-23  & 4&  7.3 & 2.9& 0.25& 5.9  & 0.6& 0.23\\
HE\,1152-0355 & 5&  8.2 & 1.7& 0.25& 7.9  & 0.5& 0.22\\
HE\,1305+0007 & 5&  7.65& 2.4& 0.31& 7.35 & 1.2& 0.23\\
HD\,187861    &10&  7.6 & 2.3& 0.32& 7.5  & 1.3& 0.31\\
HD\,5223      &10&  7.2 & 1.4& 0.24& 7.2  & 0.5& 0.21\\
HD\,122563    &15&  5.8 & 0.8& 0.21& 6.8  & 0.9& 0.21\\ \hline
 & & & & & & & \\
\colhead{Star} &\colhead{}&\colhead{$\log\epsilon$(F)}&\colhead{[F/Fe]}&\colhead{$\sigma(\log \epsilon$(F))}&\colhead{$\log\epsilon$(Na)}&
\colhead {[Na/Fe]}&\colhead{$\sigma(\log \epsilon$(Na))}\\ \hline
CS\,22942-19  &&$<$4.2  &$<$ 2.1&\nodata &   5.1&  1.3 &0.12\\
CS\,29497-34  &&$<$3.5  &$<$ 1.8&\nodata &   4.5&  1.1 &0.12\\
CS\,30314-67  &&$<$2.0  &$<$ 0.3&\nodata &   3.5& 0.05 &0.13\\
CS\,22948-27  &&$<$3.6  &$<$ 1.5&\nodata &   4.3&  0.5 &0.14\\
CS\,29502-92  &&$<$3.9  &$<$ 2.5&\nodata &$<$4.0&$<$0.9 &\nodata\\
CS\,30322-23  &&$<$1.8  &$<$ 0.6&\nodata &$<$3.8&$<$0.9 &\nodata\\
HE\,1152-0355 && ~3.90  &  ~0.64& 0.30  &\nodata&  \nodata&\nodata\\
HE\,1305+0007 &&$<$3.5  &$<$ 1.4&\nodata &   4.0&  0.2 &0.16\\
HD\,187861    &&$<$3.7  &$<$ 1.6&\nodata &$<$3.0&$<$-0.8 &\nodata\\
HD\,5223      &&  ~4.00 &  ~1.44& 0.24   &   4.1& -0.2 &0.12\\
HD\,122563    &&$<$2.5  &$<$ 0.8&\nodata &$<$3.5&$<$0&\nodata\\ 
\enddata
 \tablecomments{The adopted solar abundances are from \citet{grevesse07} for C,N, O, F and Na, and 
\citet{grevesse98} for Ba.}
\end{deluxetable}






\end{document}